\documentclass[10pt,twocolumn,letterpaper,twocolumn]{article}
\usepackage{ol2}
\usepackage[draft]{hyperref}
\usepackage{amsmath,amsfonts,amssymb}
\usepackage{calc}
\usepackage{multirow}
\usepackage{graphicx}
\usepackage{bm}
\usepackage{color}
\usepackage{epsfig}
\usepackage{ifthen}

\begin{document}

\twocolumn[

\title{Tunable bandwidth optical rotator}

\author{Emiliya Dimova$^{1}$, Andon Rangelov$^{2,3,*}$ and Elica Kyoseva$^{3}$}

\address{
$^1$Institute of Solid State Physics, Bulgarian Academy of Sciences, 72 Tsarigradsko chauss\'{e}e, 1784 Sofia, Bulgaria\\
$^2$Department of Physics, Sofia University, James Bourchier 5 blvd., 1164 Sofia, Bulgaria\\
$^3$Engineering Product Development, Singapore University of Technology and Design, 20 Dover Drive, 138682 Singapore\\
$^*$Corresponding author: rangelov@phys.uni-sofia.bg}

\begin{abstract}
We propose and experimentally demonstrate a novel type of polarization rotator which is capable of rotating the polarization plane of a linearly polarized light at any desired angle in either broad or narrow spectral bandwidth. The rotator comprises an array of standard half-wave plates rotated at specific angles with respect to their fast-polarization axes. The performance of the rotator depends on the number of individual half-wave plates and in this paper we experimentally investigate the performance of two composite rotators comprising six and ten half-wave plates.
\end{abstract}


]


\section{Introduction}


Devices that rotate the polarization plane of a linearly polarized light at a selected angle are key elements of polarization state manipulation. Typically, these devices employ either magneto-optic effect (i.e., Faraday rotator), birefringent effect or total internal reflection \cite{Hecht,Wolf,Goldstein,Azzam,Duarte}.

The Faraday rotator is made of magnetoactive medium which is placed inside a
strong magnet. The magnetic field induces different refraction indexes for the left and right circular polarized light and, as a result, the plane of linear polarization is rotated \cite{Hecht,Wolf,Goldstein,Azzam,Duarte}.
However, the angle of rotation is strongly dependent on the wavelength of the light. Similarly, the polarization rotation of birefringent base instruments is substantially wavelength dependent. In contrast, Fresnel rhombs, which are designed to use total internal reflection, can operate over a wide range of wavelengths but induce optical beam displacement in the lateral direction \cite{Hecht,Wolf,Goldstein,Azzam,Duarte}.

It is an object of the present paper to overcome the main disadvantages of current state-of-the-art polarization rotation devices and to provide an optical scheme for rotating the polarization plane of linearly polarized light that is relatively simple, compact, inexpensive and easily tunable to different rotation angles. Moreover, the device can be used in either broadband or narrowband regime. Here we report an experimental demonstration of such a device, which comprises a sequence of ordinary half-wave plates revolved at specific angles with respect to their fast-polarization axes. For the broadband regime, the design is assembled according to the recent theoretical work of Rangelov and Kyoseva \cite{Rangelov}.

\section{Theory}


The Jones matrix for a retarder with a phase shift $\varphi $ rotated at an angle $\theta $ is conveniently parameterized in the left-right circular polarization (LR) basis as,
\begin{equation}
\mathbf{J}_{\theta }(\varphi )=\left[
\begin{array}{cc}
\cos \left( \varphi /2\right) & i\sin \left( \varphi /2\right) e^{2i\theta }
\\
i\sin \left( \varphi /2\right) e^{-2i\theta } & \cos \left( \varphi /2\right)%
\end{array}%
\right] ,  \label{retarder}
\end{equation}%
where half- and quarter- wave plates revolved by an angle $\theta $ are described by $\mathbf{J}_{\theta }(\pi )$ and $\mathbf{J}_{\theta }(\pi /2)$, respectively.

A sequence of $N$ wave plates, each with a phase shift $\varphi $ and rotated at an angle $\theta _{k}$, then realizes the total Jones matrix
\begin{equation}
\mathbf{J}^{\left( N\right) }=\mathbf{J}_{\theta _{N}}\left( \varphi \right)
\mathbf{J}_{\theta _{N-1}}\left( \varphi \right) ...\mathbf{J}_{\theta
_{1}}\left( \varphi \right) .  \label{Jn}
\end{equation}%
Our first goal is to design the sequence of half-wave plates such that the composite Jones matrix $\mathbf{J}^{\left( N\right) }$ from Eq. \eqref{Jn} produces a Jones matrix of a half-wave plate which is highly robust to variations in the phase shift $\varphi $ around
a selected value of $\varphi$. For the broadband composite half-wave plate we require a flat rotation profile around phase shift $\varphi =\pi $ \cite{Rangelov}, and for the narrowband composite half-wave plate we require a flat rotation profile around phase shift $\varphi =2\pi $. We achieve this through the use of the control parameters, $\theta_k$. That is, we fix one of the $N$ rotation angles $\theta_k$ such as the Jones matrix $\mathbf{J}^{\left( N\right) }$ from Eq. \eqref{Jn} is that of a half-wave plate. We use the rest $N-1$ rotation angles to set the first $\left( N-1\right) /2$ non-vanishing derivatives of $\mathbf{J}^{\left( N\right) }$ with respect to the phase shift $\varphi $ to zero at the desired value of $\varphi $, either for broadband (BB)
or narrowband (NB):
\begin{subequations}
\label{nullify}
\begin{eqnarray}
\text{BB:} &\text{ }\left[ \partial _{\varphi }^{k}\mathbf{J}%
^{\left( N\right) }\right] _{\varphi =\pi }=0\text{ }\left( k=1,2,...,\frac{%
N-1}{2}\right) ,  \label{BB} \\
\text{NB:} &\text{ }\left[ \partial _{\varphi }^{k}\mathbf{J}%
^{\left( N\right) }\right] _{\varphi =2\pi }=0\text{ }\left( k=1,2,...,\frac{%
N-1}{2}\right) .  \label{NB}
\end{eqnarray}%
\end{subequations}

In such a manner we obtain a system of $N$ coupled nonlinear algebraic equations for the $N$ phases, which have multiple solutions for both the broadband and the narrowband conditions. Several solutions for the rotation angles $\theta _{k}$ of the individual half-wave plates for three and five composite sequences are shown in Table \ref{Table1}. The table shows sequences for broadband and
narrowband half-wave plates, which we use for experimental realization of the polarization rotator.

\begin{table}[tbh]
\caption{Calculated angles (in degrees) of the optical axes of the individual half-wave plates to implement composite sequences of broadband and narrowband half-wave plates.}
\centering
\begin{tabular}{c}
\hline
\begin{tabular}{cc}
\hline
\multicolumn{2}{c}{Broadband half wave plates} \\ \hline
N & ($\theta _{1};\theta _{2};..;\theta _{N}$) \\
3 & (30; 150; 30) \\
5 & (51.0; 79.7; 147.3; 79.7; 51.0) \\ \hline
\end{tabular}
\\
\begin{tabular}{cc}
\multicolumn{2}{c}{Narrowband half wave plates} \\ \hline
N & ($\theta _{1};\theta _{2};..;\theta _{N}$) \\
3 & (150;90;30) \\
5 & (137.0;93.7;158.2;52.7;31.2) \\ \hline
\end{tabular}
\\ \hline
\end{tabular}%
\label{Table1}
\end{table}

\begin{figure*}[tbh]
\centerline{\includegraphics[width=1.1\columnwidth]{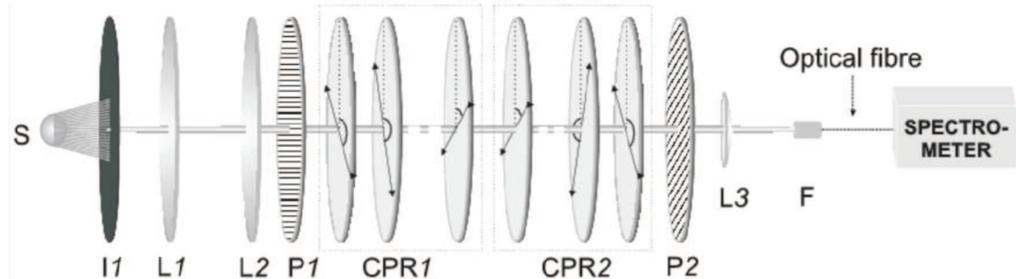}}
\caption{(Color online) Experimental setup. The source $S$, irises $I$, lens
$L_{1}$, lens $L_{2}$ and polarizer $P_{1}$ form a collimated beam of white
polarized light. Polarizer $P_{2}$ and lens $L_{3}$ focus the beam of output
light onto the entrance $F$ of an optical fibre connected to a spectrometer.
The two parts of the composite polarization rotator, which is constructed by
a stack of multiple-order half-wave plates, are denoted as CPR1 and CPR2 in
the figure.}
\label{fig1}
\end{figure*}



\section{Experiment}


Once we have the recipes for broadband and narrowband composite half-wave plates (cf. Table~\ref{Table1}) we can experimentally construct the respective polarization rotators. It is well-known  \cite{Rangelov,Kim}, that two crossed half-wave plates realize a polarization rotator, where the angle between their fast axes is half the angle of polarization rotation. Here, we extend the idea further and utilize two crossed identical broadband or narrowband composite half-wave plates. In detail, we construct the polarization rotators as a sequence of two composite half-wave plates with angles $\theta_k$ from Table \ref{Table1}, which are crossed at an angle with respect to their fast axes. Such an arrangement rotates the polarization plane of a linearly polarized light at an angle equal to twice the angle between the optical axes of the composite half-wave plates. We use both broadband and narrowband composite half-wave plates as building blocks to realize broadband and narrowband polarization rotators, respectively.

We study the experimental properties of the composite rotators by analyzing the polarization of the transmitted light. The experimental setup, shown in Fig.~\ref{fig1}, consists of three main parts: source of polarized white light, composite polarization rotator and light-analyzing part. A collimated beam of polarized light with continuous spectrum was obtained using a 10 W Halogen-Bellaphot (Osram) lamp with DC power supply, iris, two lenses and a polarizer. The iris $I$ imitates a point source of white non-polarized light, which has been placed in the focus of the first lens $L_{1}$ (f=35 mm) and additionally collimated by second lens $L_{2}$ (f=150 mm). The formed light beam is linearly polarized after passing through the first polarizer $P_{1}$ (Glan-Tayler, 210-1100 nm, borrowed from a Lambda-950 spectrometer).

The composite half-wave plates we used for the experiments comprised $k = \{3,5\}$ ordinary multi-order quarter-wave plates (WPMQ10M-780, Thorlabs), which perform as half-wave plates at $\lambda $=763 nm. Each wave plate (aperture of $1^{\prime \prime }$) was assembled onto a separate RSP1 rotation mount. The optical axes of all wave plates were determined with an accuracy of $1^{\circ }$. Each of them was rotated at the respective theoretically calculated angle $\theta _{k}$ (Table \ref{Table1}). Lastly, the wave plates were slightly tilted \cite{Peters} to reduce unwanted reflections.

The analysis of the rotation effect of the composite polarization rotator was done by a system of an analyzer $P_{2}$ (same type as $P_{1}$), a plano-convex lens $L_{3}$ ($f$=20 mm) and a two-axis micro-positioner, which were used to focus the light beam onto the optical fibre entrance $F$ connected to a grating monochromator (Model AvaSpec-3648 Fiber Optic Spectrometer with controlling software AvaSoft 7.5). With the available light source and monochromator, we obtained a reliable spectral transmission data in the range of 400-1100 nm.

The measurement procedure was similar to the one described in Ref. \cite{Dimova}. We used a single
beam spectrometer, thus all experiments started with measurement of the dark
and reference spectra. The dark spectrum, taken with the light path blocked,  is further automatically used to correct for hardware offsets. The reference spectrum is usually taken with the light source on and using a blank sample
rather than the sample under test. In our case, however, we measured the transmission spectrum of the already assembled composite polarization rotator, with the axes of the polarizer $P_{1}$, the analyzer $P_{2}$ and the
fast axis of the single wave plates all set to be parallel. The measured light spectrum was used as a reference for the subsequent measurements.

\begin{figure}[tbh]
\centerline{\includegraphics[width=1\columnwidth]{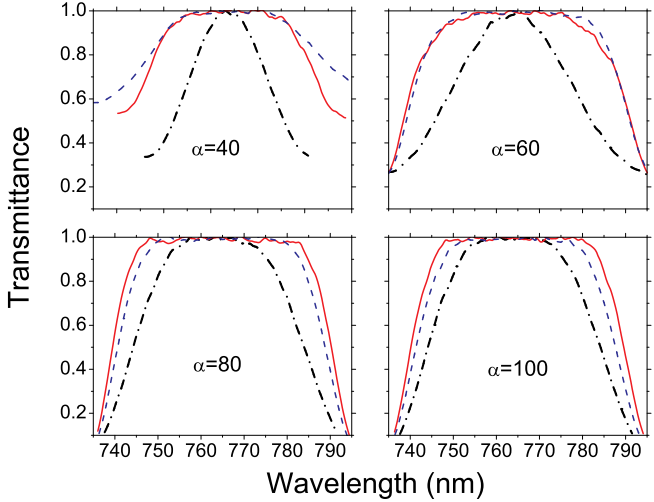}}
\caption{(Color online) Measured transmittance for two different composite
broadband rotators. The blue dash line represents a rotator with six half-wave
plates, while the red solid line represents a rotator with ten half-wave plates. The
black dash-dotted line represents rotator comprising two half-wave plates for easy
reference. }
\label{fig2}
\end{figure}


\begin{figure}[tbh]
\centerline{\includegraphics[width=1\columnwidth]{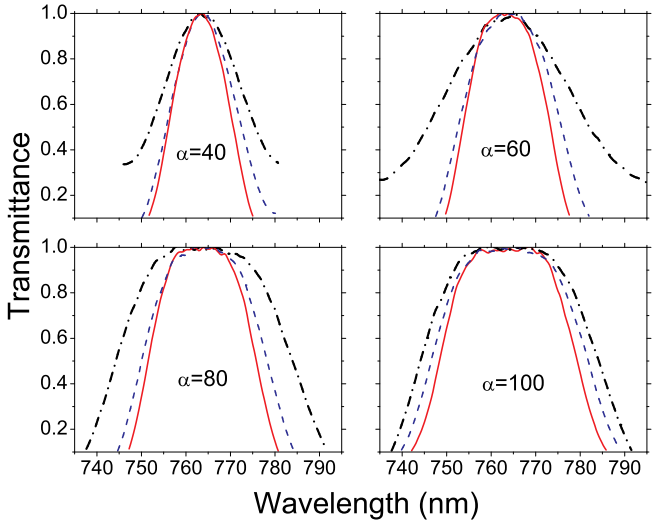}}
\caption{(Color online) Measured transmittance for two different composite
narrowband rotators. The blue dash line represents a rotator with six half-wave
plates, while the red solid line represents a rotator with ten half-wave plates. The
black dash-dotted line represents rotator comprising two half-wave plates for easy
reference. }
\label{fig3}
\end{figure}


Finally, we assembled the two composite half-wave plates and rotated the first one at an angle $\alpha /4$ clockwise and the second composite half-wave plate at an angle $\alpha /4$ anticlockwise. The analyzer $P_{2}$ was set to $\alpha $ with respect to $P_{1}$. The transmission spectrum of the realized composite polarization rotator was recorded and scaled to the reference spectrum. The unavoidable losses due to reflections and absorptions from any single wave plate were thereby taken into account.

We show the polarization rotation by the broadband and narrowband composite polarization rotators at four different rotation angles ($\alpha =40^{0},60^{0},80^{0},100^{0}$) on Figs. \ref{fig2} and \ref{fig3}, respectively. For comparison and easy reference of the broadband and narrowband behavior of the composite polarization rotators, we include the measured transmittance of a polarization rotator constructed of two ordinary half-wave plates. As predicted, we observe that the broadband and narrowband composite polarization rotators outperform the ordinary polarization rotator, and that the broadening and narrowing of the bandwidth increases with the number of half-wave plates in the sequence.

Next we turn our attention to the problem of misalignment of the orientation angles of the optic axis for each wave plate in the array; as such errors are always present in an experimental setup. For the case of broadband rotators, such errors are compensated automatically and do not affect the spectra visibly. The last statement we see by introducing additional misalignment and realising that the overall results in Fig.~\ref{fig2} are not changed. But for narrowband rotators the precise control of the orientation angles is more crucial, therefore a better control of the orientation angles will lead to considerable improvement in the narrowing of our spectrum in Fig.~\ref{fig3}.


\section{Conclusion}


In this paper, we demonstrated a novel, simple, compact, and inexpensive composite polarization rotator which is capable of rotating the polarization plane of linearly polarized light at any desired angle. We showed that the bandwidth of the composite polarization rotator can be freely controlled by varying the number of the constituent half-wave plates. Furthermore, we demonstrated that the constructed polarization rotator can be tuned to operate in both broadband and narrowband regime by appropriately adjusting the rotation angles of each of the half-wave plates.

We note the universal principle that two crossed half-wave plates serve as a polarization rotator, and therefore any previously suggested achromatic half-wave plates \cite{Pancharatnam1,Pancharatnam2,Koester,Title,McIntyre,Ardavan} may be
used to achieve the case of broadband polarization rotator.

We acknowledge financial support by Singapore University of Technology and Design Start-Up Research Grant, Project No. SRG EPD 2012 029 and SUTD-MIT International Design Centre (IDC) Grant, Project No. IDG 31300102.
We are grateful to Georgi Popkirov for valuable discussions and suggestions. 

\end{document}